# A fundamental correlative spectroscopic study on Li$_x$NiO$_2$ and NaNiO$_2$

Quentin Jacquet[1*], Nataliia Mozhzhukhina[2], Peter N.O. Gillespie[3], Gilles Wittmann[4], Lucia Perez Ramirez[4], Federico G. Capone[5,6], Jean-Pascal Rueff[4], Stephanie Belin[4], Rémi Dedryvère[7,6], Lorenzo Stievano[8,6], Aleksandar Matic[2], Emmanuelle Suard[9], Nicholas B. Brookes[10], Alessandro Longo[10,11], Deborah Prezzi[3], Sandrine Lyonnard[1*] and Antonella Iadecola[5,6*]

[1] Univ. Grenoble, Alpes, CEA, CNRS, IRIG, SyMMES, F-38000 Grenoble, France

[2] Department of Physics Chalmers University of Technology, 41296 Göteborg, Sweden

[3] Nanoscience Institute of the National Research Council (CNR-NANO), 41125 Modena, Italy

[4] Synchrotron SOLEIL, L'Orme des Merisiers, Saint Aubin BP 48, 91190 Saint Aubin, France

[5] Sorbonne Université, CNRS, Physicochimie des Électrolytes et Nanosystèmes Interfaciaux, F-75005 Paris, France

[6] Réseau sur le Stockage Electrochimique de l'Energie (RS2E), CNRS FR 3459, France

[7] IPREM, CNRS, Université de Pau & Pays Adour, E2S-UPPA, 64000 Pau, France

[8] ICGM, Univ. Montpellier, CNRS, ENSCM, 34100 Montpellier, France

[9] Institut Laue-Langevin, 71 avenue des Martyrs, CS 20156, 38042 Grenoble cedex 9, France

[10] European Synchrotron Radiation Facility, F-38043 Grenoble Cedex, France

[11] Istituto per lo Studio dei Materiali Nanostrutturati (ISMN)-CNR, UOS Palermo, 90146 Palermo, Italy

*Corresponding authors: antonella.iadecola@synchrotron-soleil.fr , sandrine.lyonnard@cea.fr, quentin.jacquet@cea.fr

**Abstract**

The intimate correlation between the local atomic arrangement and electronic states in Li-ion battery cathode materials plays a crucial role in determining their electrochemical properties, including capacity, cycling stability, and rate capability. Despite almost 30 years of research efforts on high performance cathodes based on Ni rich layered oxides, there is still no consensus on $LiNiO_2$ local atomic and electronic structure. Ni sites could be either Jahn-Teller distorted or bond disproportionated and the role of Ni and oxygen in the charge compensation mechanism remains unclear. In this study, we compare the local and electronic structure of $LiNiO_2$ and $NaNiO_2$, a long-range Jahn-Teller system, using a novel approach which aims at correlating the results from bulk spectroscopy techniques, particularly under operando conditions, obtained on standard samples to ensure sample interoperability and enhance the reliability and robustness of our results. Despite being a site-selective and local technique, XAS is unable to discriminate between the proposed scenarios, as confirmed also by theoretical calculations. On the contrary, Raman spectroscopy show local structural differences between monoclinic distorted $NaNiO_2$ and rhombohedral $LiNiO_2$. Additionally, HAXPES confirms the presence of multiple formal oxidation states for Ni, and RIXS data provides evidence of $3d^8$ states, confirming the negative charge transfer character of Ni and some degree of bond disproportionation in $LiNiO_2$. Regarding the charge compensation mechanism, XRS and RIXS support the participation of oxygen holes in the redox activity, while Raman spectroscopy does not detect molecular oxygen. By combing several high-fidelity spectroscopy datasets, this study shows the value of correlative characterization workflows to provide insights into complex structural-electrochemical relationships.

**Introduction**

Immediately after the introduction of rock-salt layered $LiCoO_2$ as positive electrode material in secondary Li-ion batteries, its nickel-containing counterpart, $LiNiO_2$ (LNO) was considered as the next promising candidate. Indeed, LNO features high theoretical capacity close to 275 mAh/g and high working potential attributed to $Ni^{4+}/Ni^{3+}$ redox couple. However, charge/discharge cycling at high voltage was observed to lead to substantial performance decay. Substitution of Ni with Al, Co or Mn increased cycling stability at the expense of energy density, leading the commercialisation of NCA - $LiNi_xCo_yAl_zO_2$ - and NMC - $LiNi_xCo_yMn_zO_2$ with x + y + z = 1. Nowadays, profiting from almost 30 years of system optimisation, the battery community is getting back to LNO to further increase the energy density, proposing new stabilisation strategies including electrolyte optimisation, particle size and monolithic crystals, doping and particle coating[1,2].

Despite the tremendous interest and effort, the atomic and electronic structure of pristine LNO as well as the charge compensation mechanism occurring during its delithiation are still under debate. Regarding the electronic structure, the debate stems from the role of nickel and oxygen in the charge compensation mechanism. Conventional ionic distribution of charges implies the existence of $Li^+$, $Ni^{3+}$ ($d^7$) and $O^{2-}$ ions, with $Ni^{3+}$ oxidizing to $Ni^{4+}$ when 1 $Li^+$ is extracted from the cathode. However, recent theoretical studies reported that the effective Ni charge in LNO and $NiO_2$ is actually closer to that in NiO, which contains $Ni^{2+}$ ($d^8$), and that most of the charge compensation occurs at the $O^{2-}$ site which oxidizes to $O^-$ [3]. This reminds of earlier studies of rare earth nickelates, carried out by the physics community in which Ni – formally $Ni^{3+}$ if considering a ionic charge distribution – was actually found to be $d^8\underline{L}$ (where $\underline{L}$ denoted a ligand hole), which disproportionates into $d^8$ and $d^8\underline{L}^2$ at low temperature[4,5]. This brings up a second debate regarding LNO, which pertains then local atomic distribution around the Ni centres. LNO, NaNiO$_2$ (NNO) and AgNiO$_2$ (ANO) are isoelectronic layered oxides and all feature two sets of Ni-O bond lengths (short and long). For NNO and ANO, short and long Ni-O bonds are long ranged ordered into cooperative Jahn-Teller (JT) distortion (Ni$^{3+}$, 3d$^7$) for NNO[6], whereas ANO features a long range ordered bond disproportionation (BD) (3Ni$^{3+}$ → Ni$^{2+}$ and 2 Ni$^{3.5+}$)[7–9]. Conversely, LNO does not show any long range ordering observable by diffraction, hence the question regarding the distribution of long and short Ni-O bonds previously revealed by EXAFS remains open[10,11]. In particular, it is not clear whether the short and long distances appear together in all octahedra, as in JT distorted systems, or whether they belong to different octahedra, as in bond disproportionated (BD) systems.

Several theories have been proposed so far regarding the electronic and atomic structure of LNO. In 2005, Chung *et al.* proposed a static random orientation of JT distortion based on Neutron Pair distribution function (nPDF)[12]. More recently, Sicolo *et al.* presented a random dynamic reorientation of these JT octahedra at the ps time scale[13]. Genreith-Schriever and coworkers suggested a mixed-phase regime where JT distorted and undistorted domains coexist, and antisite defects pin the undistorted domains at low temperatures, impeding cooperative ordering at a longer length scale[14]. On the other hand, Chen *et al.* reported similar formation energies for JT and BD Ni$^{3+}$ systems, suggesting their possible coexistence in the material.[15] Moreover, Foyevtsova *et al.* proposed LNO to be a charge disproportionated negative charge transfer material with a 3d$^8$ + 3d$^8\underline{L}$ + 3d$^8\underline{L}^2$ electronic configuration ($\underline{L}$ corresponding to a ligand hole in the O 2p orbitals) [16,17]. Concerning the redox process upon delithiation, some studies point on the increase of Ni-O bond covalency - with the electron-hole density shifting towards O when Ni is highly oxidized [16,18] - while others argue that anionic redox activity is triggered [3,19].

In the light of these different proposed theories, experimental evidences are needed to lift the controversies for LNO, and especially to answer the following open questions: i) does the atomic structure result from JT distortion and/or BD mechanism? ii) is the electronic structure of pristine LNO described as a 3d$^7$, 3d$^8\underline{L}$, 3d$^8$ + 3d$^8\underline{L}$ + 3d$^8\underline{L}^2$, or as a mixture of all these possible states? iii) what is the charge compensation mechanism upon cycling? To answer to these questions, new experimental data possessing two important properties are needed: (1) advanced spectroscopies giving average bulk information, as it is well known that Ni surface state in Ni-rich cathode materials may be substantially different from the bulk; (2) standard reproducible LNO samples, with a controlled concentration of Li$_{Ni}$ and Ni$_{Li}$ antisite defects.

In this work, we applied a correlative bulk-spectroscopy workflow to investigate/characterize LNO samples prepared following a standardized workflow. This approach was carried out within the framework of the European project BIG-MAP, where partners have shared the same pristine or cycled samples for laboratory and large scale facility studies (big-map.eu). *Operando* X-ray absorption spectroscopy (XAS) and Raman spectroscopy, as well as *ex situ* hard X-ray photoelectron spectroscopy (HAXPES), X-ray Raman scattering (XRS), resonant inelastic X-ray scattering (RIXS) are performed on the same pristine and cycled commercial LNO polycrystalline electrodes (Figure 1). Theoretical calculations were also carried out to help the interpretation of selected experimental datasets. The results of these investigations are discussed together to provide a unified answer to the previously mentioned debated questions on local atomic and electronic structure of LNO.

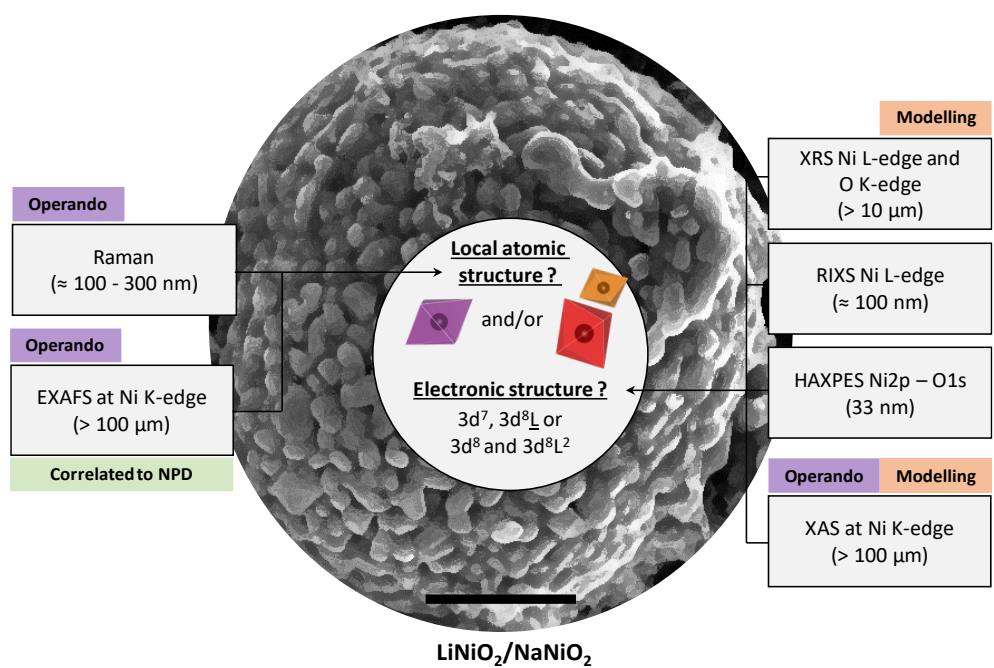

**Figure 1:** Schematic representation of the spectroscopy workflow. Filtered SEM image of a LNO particle is shown at the background (scale bar is 1 µm) together with the different techniques used in this work (and their penetration depth under brackets).

Results:

Before addressing the questions on the electronic and local atomic structure, the long range crystallographic structure of LNO was determined by Rietveld refinement of *ex situ* high resolution and *operando* neutron diffraction patterns (Figure S1, S2, S3). The LNO studied in this work is phase pure, with 2.3(2) % of Ni in the Li layer and 97.7(2)% Ni in the Ni layer. During delithiation, the expected H1, M, H2 and H3 phase sequence is observed with cell parameter evolution in agreement with previous reports [20]. Charged sample features 3.4(6)% Ni in the Li layered and 96.5(6)% Ni in the Ni layer.

*Operando* XAS at Ni K-edge is a suitable tool to acquire bulk information on the Ni local environment and oxidation state upon cycling. The LNO electrode was cycled using a homemade electrochemical cell equipped with X-ray transparent Be windows, showing performance comparable to coin cells and described elsewhere. The possible occurrence of beam damage was assessed by measuring different positions at specific state of charges during cycling. The edge position of the *operando* XANES spectra (Figure 2a) shows a progressively shift of the edge position towards higher energies upon charge, in agreement with previous studies. This blue shift is usually interpreted in terms of $Ni^{4+}/Ni^{3+}$ redox activity [21]. It is important to recall that, while the edge position is often used to determine the formal oxidation state of the Ni ions, this assumption implies a fully ionic character of the Ni-O bonds, which is not accurate for describing strongly covalent Ni-O bonds[3]. The XAS dataset was then analysed using a chemometric approach including Principal Component Analysis (PCA) and Multivariate-Curve Resolution Alternating Least-Squares (MCR-ALS): three independent spectral components are necessary to describe the evolution of the XAS spectra upon cycling (Figure 2b and Figure S4). The first and third component (PC1 and PC3) correspond to the XAS spectra of the electrode at OCV and 4.3 V, respectively, whereas the second (PC2) has a maximal concentration of 50% at the middle of charge. The EXAFS analysis was conducted directly on the MCR-ALS pure components to obtain quantitative information on the local radial distribution of the neighbours around the absorber Ni atoms (Figure 2b). The main peaks in the Fourier Transforms correspond to the Ni-O and Ni-Ni bonds. In pristine LNO, different structural models were used to fit the Ni-O first shell (see SI, Table S1). The best fits are obtained with 4 short (1.92 Å) and 2 long (2.04 Å) Ni-O bonds, as well as with a model using 2+2+2 Ni-O distances (1.87, 1.92 and 2.04 Å), even though reasonable fits are also obtained for 6 equivalent oxygens distributions. For the Ni-Ni bond, the best fit requires only one distance, corresponding to 6 Ni atoms at 2.88 Å. Upon delithiation, the distorted $NiO_6$ octahedra are found to progressively convert into a regular one (Table S2 in SI). Indeed, the fit of the PC2 EXAFS signal results in 5.3 Ni-O bonds at 1.89 Å and only 0.7 Ni-O at 2.08 Å, whereas the Ni-$O_6$ polyhedra are undistorted in PC3, with 6 equivalent oxygen atoms at 1.87 Å. In parallel, the Ni-Ni distance shrinks to 2.83 Å in PC3. It is important to note that the monoclinic distortion indexed in the C2/m space group and reported for 0.8<x<0.4 in $Li_xNiO_2$ is not observed by EXAFS. Indeed, while the Ni-O distance distribution obtained by *operando* neutron diffraction increases across the monoclinic transition [22], the EXAFS data reveal a progressive decrease of the Ni-O distance distribution (Figure 2c). This shows that, even during delithiation, the long-range description of the structure diverges from the local atomic one. To try to assess the possible presence of JT distortions or BD in LNO, the Ni local environment in LNO was compared to that in $NaNiO_2$ [23]. By refining the EXAFS spectrum of NNO (Figure 2d), we found that both $NiO_6$ octahedra and Ni-Ni environment are distorted, differently from the equivalent Ni-Ni bonds in the second coordination shell of LNO [24]. Moreover, 4 Ni-O short distances close to 1.92 Å and 2 Ni-O long distances at 2.15 Å are needed to fit the first shell in NNO, the latter distance being significantly longer that in LNO. Even though distortions are clearly less significant in LNO than in NNO, the EXAFS analysis is unable to rigorously discriminate between the JT or BD scenario. Some insights could come from the simulation of their respective XANES spectra.

The *ab initio* Ni K-edge XANES spectra of pristine LNO and NNO (Figure 2d) were calculated from their potential crystal structures by using the XSpectra code [25,26]. For NNO, we used the established collinear-JT structure (space group C2/m). For LNO, both JT and BD structures were used [16,27]. There are shape differences between simulated LNO and NNO, notably an increased intensity at 8345 eV. This difference is also observed in the experimental XANES spectra (Figure 2d – right panel). However, only very subtle differences are visible when comparing the simulated XANES spectra of JT and BD pristine LNO structures (Figure 2d – middle panel). We can therefore conclude that the influence of JT or BD in LNO cannot be unambiguously addressed by XAS.

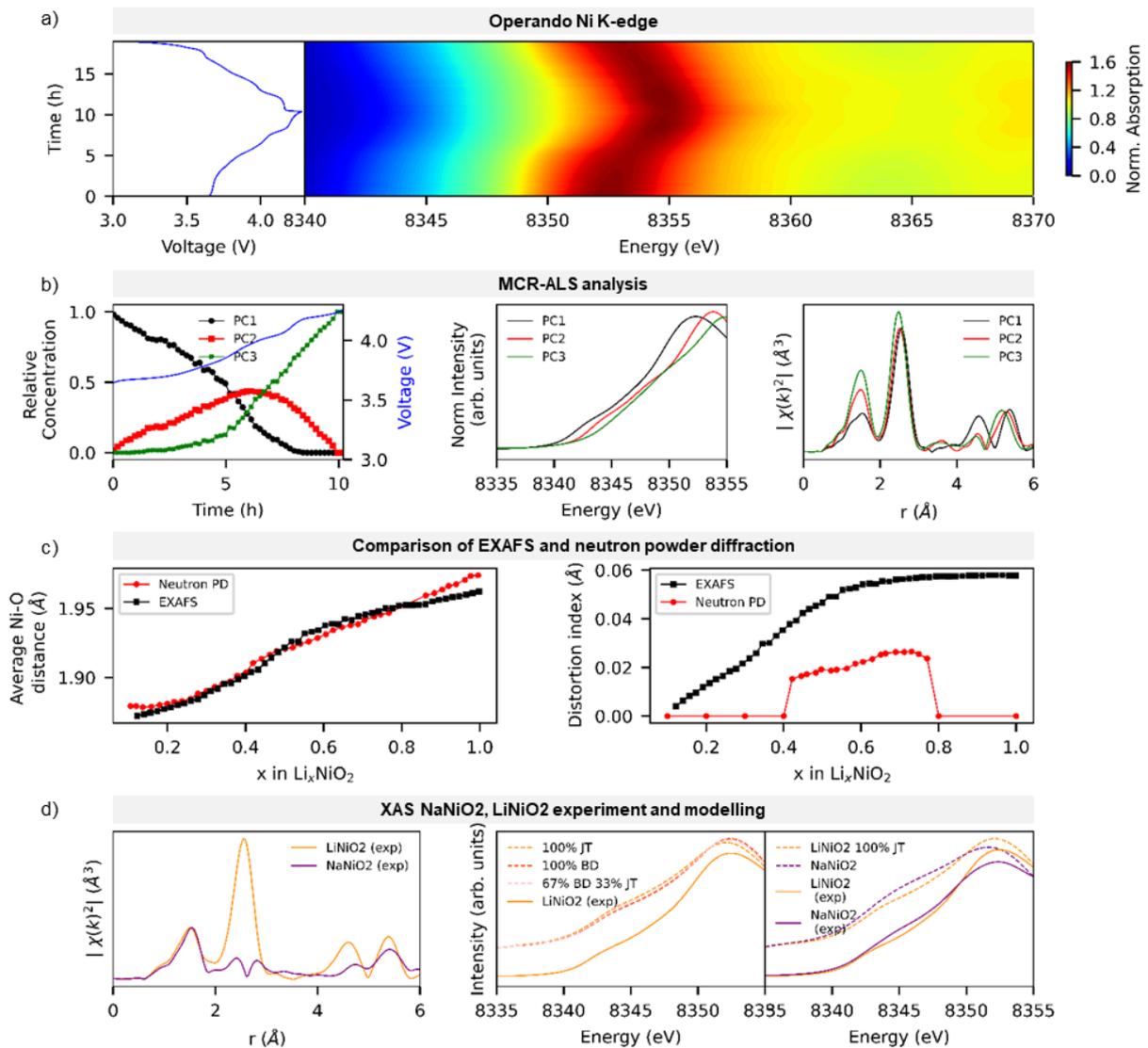

**Figure 2:** a) Operando Ni K-edge spectra evolution (right panel) together with the voltage profile (on the left panel). b) Concentration, XANES and FT of the EXAFS oscillation of the three reconstructed principal components obtained by MCR-ALS. c) Average Ni-O and distortion index obtained from operando EXAFS and neutron diffraction (left and right panel). d) Experimental FT of the EXAFS oscillations for LiNiO$_2$ and NaNiO$_2$ on the left panel. Right panel shows the experimental and calculated XANES (full and dashed line) for LiNiO$_2$ and NaNiO$_2$, in orange and purple respectively. Calculated spectra are displayed with an intensity offset for clarity.

Additional information on the local structure can be retrieved by Raman spectroscopy, which has a probing depth of approximately a few hundred nm for LNO. And provides structural information on the local short-range order. At a first glance, the Raman spectrum of LNO features two peaks at 480 and 542 cm$^{-1}$ (Figure 3ab). By comparison with LiCoO$_2$ crystallizing in the same R-3m space group, these peaks can be attributed to $E_g$ Ni-O-Ni bending mode (480 cm$^{-1}$) and $A_{1g}$ Ni-O stretching mode (542 cm$^{-1}$).[28,29,6] However, LNO bands are much broader than those of LiCoO$_2$ and very different from NaNiO$_2$ (Figure 3ab), for which $E_g$ band splitting into $A_g$ and $B_g$ is expected and three Raman active bands are predicted when moving to the C2/m space group (2$A_g$ and $B_g$)[30,31]. Thus, we can confirm that a different local environment is observed in both layered NNO and LNO. *Operando* Raman spectroscopy was used to follow the evolution of the LNO spectral signature upon cycling (Figure 3cd). Positions, intensities, and widths of both $E_g$ and $A_{1g}$ bands change reversibly during cycling. Notably, upon delithiation both bands become sharper and more intense. By fitting all spectra, a more precise description of the evolution of band positions is obtained. They change in a non-monotonic fashion between 467 and 477 cm$^{-1}$ for the $E_g$ mode, and between 544 and 553 cm$^{-1}$ for the $A_{1g}$ mode. Moreover, no splitting of the $E_g$ bands is observed by *operando* Raman spectroscopy during the monoclinic distortion occurring at 0.8<x<0.4 in Li$_x$NiO$_2$. On the contrary, broad bands for LNO are becoming narrower during charge suggesting decrease of the Ni-O bond distance distribution, in agreement with EXAFS analysis. Regarding band intensities, a similar increase is observed in the *operando* Raman spectroscopy investigation of Ni-rich layered cathodes[29,32,33]. One explanation for this trend could lie in optical skin depth effect, *e.g.*, the variation of light penetration depth/Raman intensity due to changes in the electronic conductivity of the material and/or of its magnetic permeability[34,35]: the higher the electronic conductivity and magnetic permeability, the lower the laser penetration depth and Raman intensity. For example, this effect is observed in LiCoO$_2$ which undergoes a metal-insulator transition, corresponding to a significant modification in Raman bands intensities[30,36]. However, for LNO the conductivities were reported to increase upon delithiation [37], which would result in a decreased optical skin depth as well as decreased Raman intensities, e.g., the opposite of the observed experimental results. Therefore, for the LNO another possible explanation could be the higher covalency of the Ni-O bonds upon delithiation that would result in more intense Raman bands [38]. Moreover, the overall increase in structural order, as inferred by the EXAFS analysis, is expected to result in more intense and sharper Raman peaks. To the best of our knowledge, no theoretical calculations of the Raman spectra of LNO provide an unambiguous reproduction of the observed features.

In order to address on the oxygen redox in LNO, the *ex situ* Raman spectrum of LNO cathode charged to 4.3 V (Figure S5) was recorded in an extended wavenumber range. Raman spectroscopy is particularly sensitive to reduced oxygen species, with peroxide stretching band appearing in interval of 700-900 cm$^{-1}$, superoxide at 1050-1200 cm$^{-1}$ and molecular oxygen at 1550-1560 cm$^{-1}$ [39,40]. Even though the presence of molecular O$_2$ was inferred by O K-edge mRIXS [19,41], we were unable to detect any peaks in the mentioned regions by Raman- e.g. no oxygen reduced species were detected in charged LNO to 4.3 V - while the probing depth of both mRIXS and Raman spectroscopy is very similar (few hundreds of nm), and Raman spectroscopy has higher sensitivity than mRIXS.

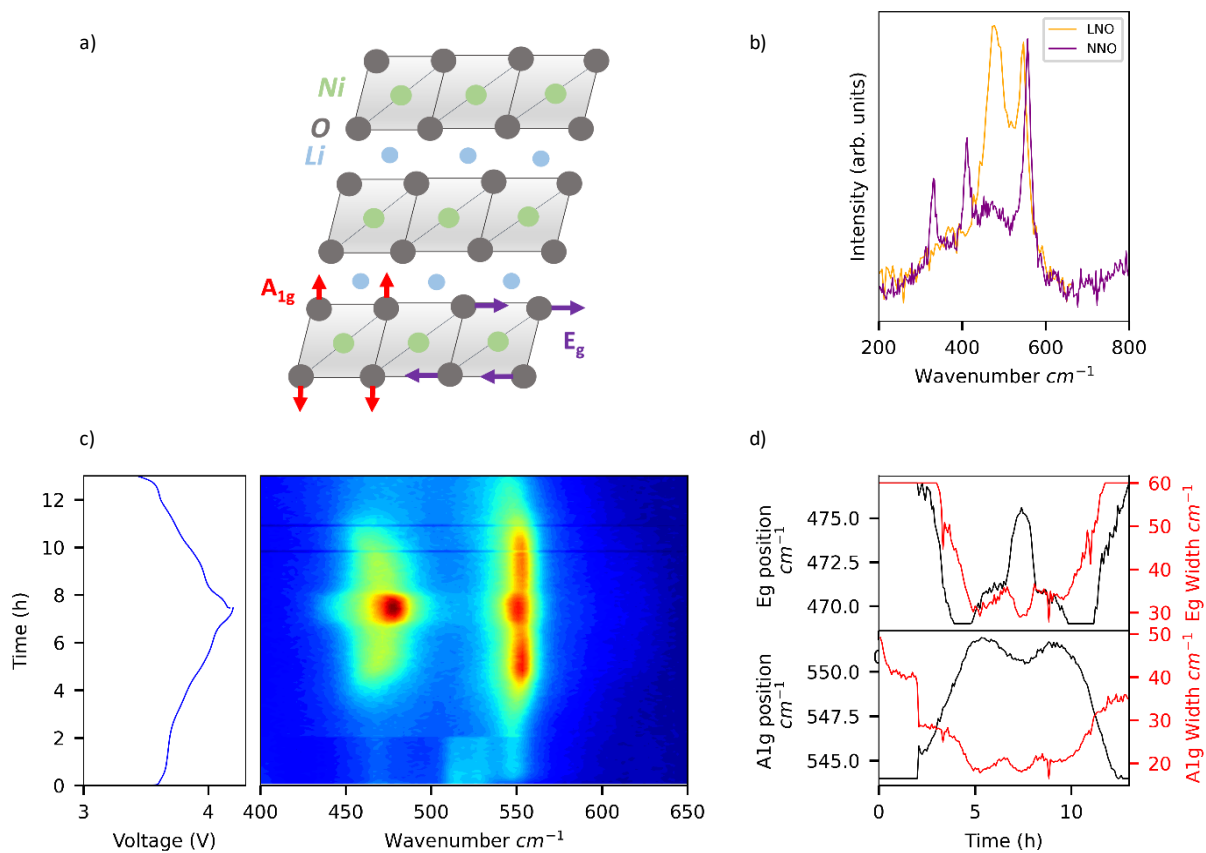

**Figure 3:** a) Schematic description of the LNO atomic structure showing the $E_g$ and $A_{1g}$ modes. b) Raman spectra of pristine LNO and NNO. c) *Operando* Raman of LNO with the voltage profile and Raman spectra evolution shown in the left and right panels, respectively. Color map represents the intensity of the Raman spectra which increases with the voltage. d) $E_g$ and $A_{1g}$ (top and bottom) peak positions and widths (black and red) obtained from fitting the *operando* dataset.

Further insights on the electronic structure of $Li_xNiO_2$ are provided by *ex situ* hard X-ray photoelectron spectroscopy (HAXPES), with a calculated probing depth of 33 nm at an incident photon energy of 8.6 keV [42]. Ni $2p_{3/2}$ and $2p_{1/2}$ HAXPES spectra of $LiNiO_2$ and $NaNiO_2$ are shown in Figure 4a. Ni 2p spectra are very sensitive to the oxidation state and local environment of nickel, as already shown in previous studies [43,44] and different shapes are observed for the main peaks and their satellites for the two materials. In particular, LNO displays a $2p_{3/2}$ satellite at 861 eV (and its parent $2p_{1/2}$ satellite at 878 eV) which is characteristic of $Ni^{2+}$ and not observed for NNO (despite the presence of 6 % of NiO impurity in NNO)[45]. In order to investigate deeper the distribution of different nickel environments in LNO, the Ni 2p spectra of pristine and charged LNO were fitted with three components corresponding to different Ni environments (Figure 4b). NNO was chosen as a reference for $Ni^{3+}$ (ionic description), while a Li-rich NMC material ($Li_{1.2}Ni_{0.13}Mn_{0.54}Co_{0.13}O_2$) after discharge at 2 V vs. $Li^+/Li$ was taken as a reference for $Ni^{2+}$ (ionic description)[43]. The third component corresponds to Ni in a more oxidized environment and is deduced from charged LNO spectra subtracted from $Ni^{2+}$ (green) and $Ni^{3+}$ (magenta) small contributions probably arising due to surface degradation (clearly observed looking at the shoulder at ∼ 855 eV in the Ni $2p_{3/2}$ spectrum) [46]. The Ni 2p spectrum of pristine LNO was then fitted with these three $Ni^{2+}$, $Ni^{3+}$ and $Ni^{4+}$ components, and their relative contributions are indicated in Figure 4b. Given the error of this approach, the relative concentration of $Ni^{2+}/Ni^{3+}/Ni^{4+}$ can be estimated roughly at ∼ 33/33/33% in pristine LNO. Note that it is not a surface effect, as shown by the comparison of Ni 2p and O 1s spectra of pristine LNO obtained at 3.7 and 8.6 keV photon energies (Figure S6): the increase of probing depth multiplies by two the bulk-to-surface intensity ratio of oxygen species in the O 1s spectrum, while the Ni 2p spectrum is not significantly modified. Unfortunately, O 1s spectra were not exploitable due to surface degradation already in the pristine electrode, thus no information on the possible anionic redox activity can be extracted from HAXPES[47].

At this stage we can conclude that *ex situ* HAXPES results confirmed the hypothesis of a bond disproportionated scenario with a mixture of $Ni^{2+}/Ni^{3+}/Ni^{4+}$ oxidation state in $LiNiO_2$ primary particles, in agreement with the operando Raman results.

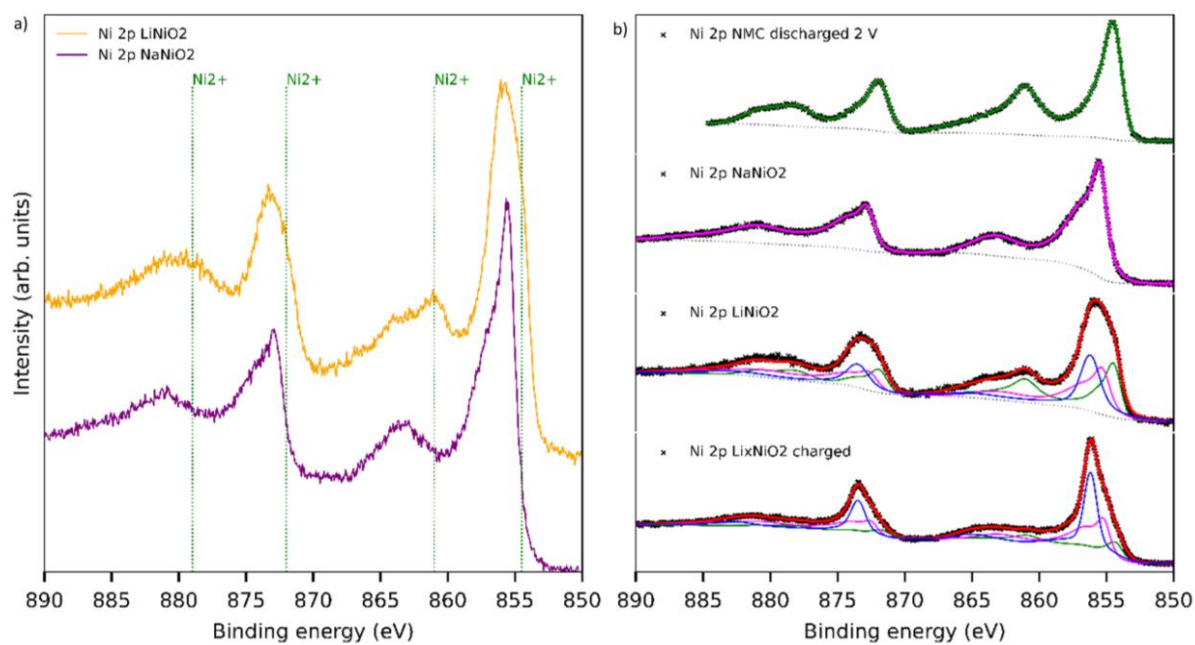

**Figure 4:** *Ex situ* Ni 2p HAXPES spectra recorded at a photon energy of 8.6 keV. a) Comparison of $NaNiO_2$ and $LiNiO_2$, the vertical green dashed lines correspond to $Ni^{2+}$ features. b) Curve fitting of the pristine and charged $LiNiO_2$ spectra using three components: discharged NMC spectra ($Ni^{2+}$ in ionic description - green), $NaNiO_2$ spectra ($Ni^{3+}$ ionic description - magenta), and a third component (blue) corresponding to $Ni^{4+}$.

*Ex situ* X-ray Raman scattering (XRS) spectra were collected at Ni $L_{2,3}$-edge and O K-edge to retrieve bulk information on the electronic structure of $Li_xNiO_2$ [48]. In dipole approximation, Ni $L_{2,3}$-edge represents the electron transition from filled Ni 2p to empty Ni 3d states, which are hybridized with ligand O 2p states; similarly, O K-edge corresponds to the 1s –>2p transition, with the pre-edge intensity arising from the hybridization of O 2p and Ni 3d states, mirroring what observed at Ni $L_{2,3}$-edge. X-ray absorption spectroscopy of Ni $L_{2,3}$-edge and O K-edge, being relatively low energy excitations, have typical probing depths ranging from a few nm to 100 nm. However, in XRS the incident energy is around 10 keV, making this technique unique to probe low energy excitation with bulk penetration depth (> 10 μm). Figure 5 shows the XRS spectra for pristine LNO, pristine NNO as well as charged $Li_xNiO_2$ (4.3 V) obtained at high q and medium q value for Ni and O edge, respectively. Generally, Ni $L_3$-edge XAS spectra of $Li_xNiO_2$ feature two peaks at 853 and 855.2 eV (peaks A and B, respectively)[49], and their relative intensity depends on the Li content, in agreement with previous reports [50]. Significantly different is the shape of the Ni XRS spectra obtained at high q value, where peak A is strongly damped in the charged LNO. The same shape is observed for Ni XRS spectra at low q value, thus in a pure dipole approximation, confirming the robustness of our results in both pristine and charged state (Figure S7, S8). Looking at the O XRS spectra, the pre-edge intensity is observed to increase during delithiation, while the main edge shifts to higher energy. A similar trend is observed in O K-edge sXAS[50], while the XRS pre-peak intensity is doubled in the charged LNO electrode compared to the pristine one. Overall, while the observed spectral features are generally consistent with the literature, two major differences should be highlighted for charged LNO: i) the peak A is strongly attenuated in the Ni $L_3$-edge spectra, and ii) the large increase of the pre-edge intensity at the O K-edge.

Seeking to assess the electronic bulk structure of LNO, we compared the experimental XRS spectra at Ni $L_3$-edge and O K-edge with calculations performed using Hilbert++ and FDMNES respectively[51,52]. The Hilbert++ method considering multiplet interactions was employed to better describe the features in the Ni $L_{2,3}$ XRS spectra. The long range structural model with rhombohedral symmetry was considered for pristine and charged LNO (with shrinkage of unit cell due to delithiation). Both peak positions and intensities are well reproduced considering the $Ni^{3+}$ and $Ni^{4+}$ formal oxidation state in pristine and charged electrode, respectively. The results at Ni $L_{2,3}$ confirm that XRS, like XAS, is not sensitive to the long range structure and it is unable to distinguish unambiguously between JT and BD scenario. A more detailed theoretical study is necessary to assess on the presence of ligand holes from Ni$L_{2,3}$-edge XRS data. FDMNES calculations at O K-edge, carried out using the same input structure of the Hilbert++ calculations, show that an increased overlapping (dilatorb) between the Ni 3d and O 2p orbitals is necessary to reproduce the intense pre-peak in charged LNO, proving the higher covalency of the Ni-O bond. Moreover, the participation of oxygen holes in the charge compensation mechanisms is evidenced by the Ni-O charge redistribution related to the different screening of oxygen atoms.

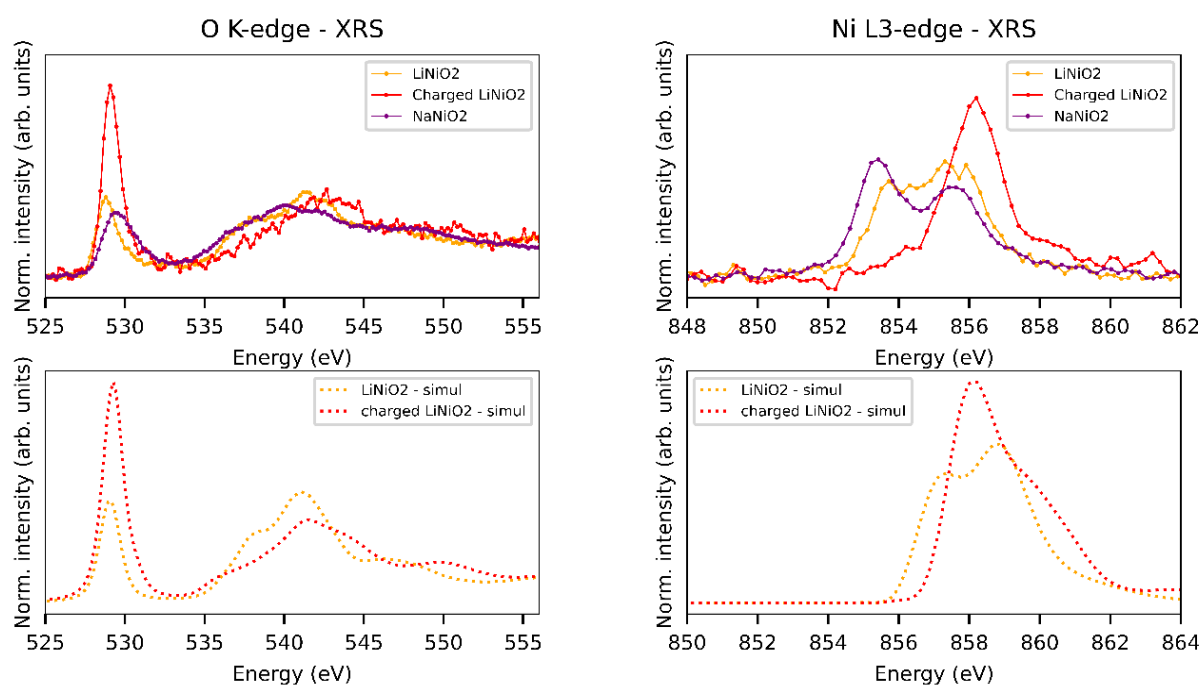

**Figure 5:** *Ex situ* XRS experimental (top) and calculated (bottom) spectra at O K-edge and Ni L-edge (left and right panel). Experimental spectra were measured for pristine LNO and NNO, and charged

**LNO. Simulations are shown for pristine and delithiated LNO. Note that O K-edge simulations are performed with FDMNES while Ni L-edge spectra are calculated using Hilbert++.**

Resonant Inelastic X-ray scattering (RIXS) at the energy corresponding to peaks A (852.5 eV) can efficiently discriminate between the possible $3d^6$, $3d^7$, $3d^8$, $3d^8\underline{L}$, $3d^8\underline{L}^2$ contributions and has been used in the rare-earth nickelate community to study the nature electronic excitation (Figure 6 a)[4]. Aside the elastic peak, the RIXS spectrum is divided in two regions: between the elastic peak and 4 eV, it is dominated by the d-d electronic transitions, and for energy loss higher than 4 eV, the main contributions are the charge transfer excitations. The energy of d-d transitions directly reflects the 3d occupancy and the crystal field intensity, making them a fingerprint of the electronic state and local structure [53,54]. Emission spectra obtained exciting peaks A are compared for LNO, NNO and simulated spectra for undistorted $3d^8$ Ni (using previously reported parameter for NiO RIXS simulations[53] with larger crystal field due to shorter Ni-O bonds in LNO compared to NiO). LNO and NNO spectra are quite different; in particular, the 1.2 eV peak is broader for NNO, in agreement with more distorted Ni environments, as observed by EXAFS. A good match between $3d^8$ simulated d-d transition energies (and hence NiO spectra) and observed transitions in LNO strongly suggests to presence of undistorted $3d^8$ in this material. Note that simulated d-d transitions energies for $3d^7$ performed by Bisogni et al.[4] feature an intense peak at 0.3 eV which is not observed in LNO or NNO spectra even varying the excitation energy, calling for further analysis on the nature of the JT transition in NNO and/or improvements in $3d^7$ modelling. Coming back to LNO, the experimental peaks are significantly broader than the simulated ones especially at high energy loss. The origin of this broadening can be understood by comparing spectra obtained during the delithiation of LNO (Figure 6 b): a strong decrease of the intensity is observed at -1.2, -2.1 and -3.4 eV during delithiation matching the energy of d-d transitions calculated for undistorted $3d^8$. This observation suggests that (i) d-d transitions resonating at Peak A originate from several environments: undistorted $3d^8$ and possibly $3d^8\underline{L}$ and/or $3d^8\underline{L}^2$, and that (ii) during delithiation, undistorted $3d^8$ Ni environment is oxidized first. Altogether, these results support the presence of some degree of bond disproportionation for LNO with the presence of undistorted $d^8$ Ni environments.

Additional information on the structure of LNO can be obtained by varying the temperature. Indeed, a weak monoclinic distortion (angle<90.1°) is observed in LNO at approx. 100 K[12]. Recent calculations explained this transition with the creation of fluctuating undistorted domains in the low temperature monoclinic distorted matrix as the temperature increases[14]. To assess the temperature dependence of Ni local environment (the ratio of distorted/undistorted $NiO_6$), Ni L-edge RIXS exciting peak A was collected at 25 K and compared to RT results (Figure 6c). The comparison shows at -1.2 eV peak shift to higher energy loss, probably caused by the increase of the crystal field splitting energy due to Ni-O thermal contraction. Moreover, peaks are narrower at low temperature excluding the presence of more distorted environments. The peak narrowing could be due to an increased fraction of undistorted $d^8$ Ni environments at low temperature suggesting an increase of the bond disproportionation at low temperature. It remains to be understood which would be the relationship between bond disproportionation ordering and the observed weak monoclinic distortion. Note that a weak monoclinic distortion was observed for $3R-AgNiO_2$ due to bond disproportionation of the $NiO_6$ octahedral site [7].

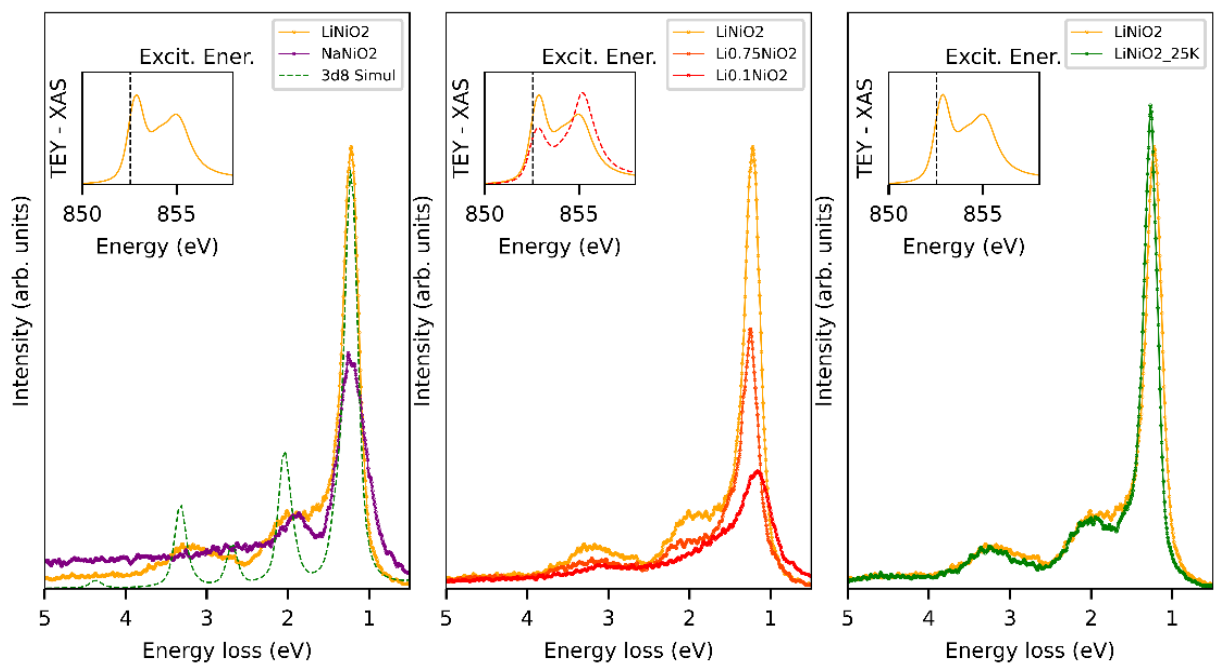

**Figure 6: RIXS spectra obtained exciting around Peak A (852.5 eV) for pristine LiNiO2 and NaNiO2 (left panel), LiNiO2, *ex situ* $Li_{0.75}NiO_2$ - 3.8 V and $Li_{0.1}NiO_2$ - 4.2 V (middle panel) and LiNiO2 – RT and 25 K (right panel). Inset showing the total electron yield (TEY) XAS spectra of LNO and $Li_{0.1}NiO_2$ along with the excitation energy used (dashed line).**

**Discussion**

In the previous section, spectroscopic datasets acquired with complementary techniques on standardised and reproducible LNO samples are reported, along with the isoelectronic sodium-containing counterpart NNO. Many of these techniques have never been applied before, to the best of our knowledge, for the characterisation of LNO, in particular *operando* Raman spectroscopy, *ex situ* Ni L-edge and O K-edge XRS and Ni L-edge RIXS or Ni 2p HAXPES on NNO. Moreover, the originality of our approach relies on the experimental workflow based on the interoperability of LNO and NNO samples, that will exclude any bias effect due to the synthesis and cycling conditions usually limiting the comparison between different datasets.

These results, taken together, can be now used to discriminate between the different possible descriptions of the local structure and electronic configuration of LNO. Ni 2p HAXPES and Ni L-edge RIXS show the presence of Ni environments similar to NiO (Ni in regular $3d^8$ configuration), which are not observed for NNO, thus indicating the occurrence of bond disproportionation in LNO. These data also show the presence of other Ni environments, including Ni in formal oxidation states +3 and +4+, without bringing, however, clear information on their local structure. No splitting of the $E_g$ band, caused by JT distortion, is observed in the Raman spectra of LNO, while it is observed for $N_{NO}$. Both the $E_g$ and $A_{1g}$ bands, however, are broad suggesting the presence of several different Ni-O environments. During delithiation, $E_g$ and $A_{1g}$ bands become sharper which is consistent with the progressive decrease of the number of the different environments. Operando XAS provided complementary information, showing a consistent narrowing of Ni-O bond length distribution even during the monoclinic transition observed by XRD or NPD between x = 0.8 and 0.4 for $Li_xNiO_2$. RIXS on $Li_{0.75}NiO_2$ also doesn't show clear sign of local distortion increase showing that the monoclinic distortion is due to long range ordering without increase in local distortion. Previously, electron diffraction unveiled Li/vacancy orderings across the monoclinic distortion of $Li_xNiO_2$ [55], proposing a larger unit cell model to account for the observed electron diffraction pattern. Of course, Li/vacancy ordering alone would not lead to the lattice distortion observed by XRD, it would likely include Ni and/or O position change [56]. Theoretical work proposed an ordering of JT distortion triggered by Li/vacancies ordering[57], but another explanation could be an ordering of the bond disproportionation with the more oxidized $NiO_6$ gathering around the ordered vacancies in the structure. In summary, our datasets and their interpretation would be in very good agreement with the charge disproportionated negative charge transfer model proposed by Foyevtsova *et al.*[16] which is also consistent with neutron PDF data from the literature[12].

Concerning the dynamic nature of LNO recently proposed by Sicolo *et al.* or Poletayev *et al.*[13,58], it should be noted that the spectroscopy techniques used in this work do not allow its observation. Indeed, the core-hole lifetime in XAS is of the order of few fs, while the dynamic of the bond ordering is three orders of magnitude higher (ps), hence giving a static average snapshot of the structure.

Regarding the charge compensation mechanism, Ni K-edge and Ni L-edge XRS do not probe directly the pure electronic density around Ni atoms, as the collected spectra are also dependent on electron removal at the oxygen sites surrounding the Ni centres. However, no sign of $d^7$ d-d transition could be observed in the RIXS spectra of LNO, in agreement with the theoretical work of Genreith-Schriever *et al.*[3] Note that RIXS was also used in the rare earth nickelate community to prove the negative charge transfer nature of these materials, which is now commonly accepted for these systems. Moreover, O K-edge XRS spectra point out on the higher covalency of the Ni-O bond upon Li extraction, which could drive to electron removal from O states, thus generating a hole charge carrier. Extended Raman spectrum on ex situ charged LNO clearly evidences the absence of molecular oxygen, despite its role on the redox activity was recently reported in high-resolution mRIXS[19].

**Conclusions**

A correlative approach based on several complementary spectroscopic methods, including hard XAS, Raman spectroscopy, HAXPES, XRS and RIXS was used, in *operando* and *ex situ* conditions, to study a set of LNO samples prepared following a standardized protocol. The results of these investigations, compared with those obtained on the Na-containing counterpart NNO, were complemented by theoretical calculations to help their interpretation. The results of these investigations suggest that LNO contains Ni in different oxidation states, in line with occurring of a charge disproportionated negative charge transfer model previously proposed in the literature rather than with a non-cooperative Jahn-Teller distortion model, which is on the contrary observed for monoclinic NNO.

Moreover, while these results provide further insights on the open questions concerning the local atomic structure and electronic structure of LNO, they have also the advantage of being recorded on the same standardised materials, thus avoiding biases which could come from different synthesis or cycling conditions. We believe that such an approach is necessary to provide a consistent basis for future studies, and in particular for more advanced interpretations obtained by theoretical approaches.

## Experimental methods

### Neutron diffraction

*Ex situ* neutron diffraction measurements were carried out at the D2B high-resolution neutron diffractometer at the Institut Laue-Langevin (ILL, Grenoble, France). Powder samples were put in cylindrical vanadium cans and measured in transmission geometry at $\lambda$ = 1.5949 Å. Due to be high amount of active material needed for neutron diffraction, ex situ charged $Li_xNiO_2$ was prepared mixing pristine $LiNiO_2$ and carbon super P, placed into a Swagelok type cell in half cell configuration (glass fiber separator, LP57 electrolyte, Li metal anode) and charged to the desired voltage at C/10. After charging, powder is washed in DMC and dried. *Operando* neutron diffraction experiment was performed on D19 (ILL, Grenoble, France). LNO powder mixed with Csp (80%wt/20%wt), and inserted into an electrochemical cell designed for neutron diffraction[59]. Two glass fibers, Li metal and deutered electrolyte (1/1 D-EC/D-DMC with 1M $LiPF_6$) are used as separators, anode and electrolyte. Neutron diffraction patterns are measured at $\lambda$ = 1.45560 Å integrating 1h of measurement time. Cycling was performed at C/35 giving 35 neutron diffraction patterns over the 1$^{st}$ charge. Rietveld refinement were performed using FullProf Suite [60]. Sensitivity analysis of operando versus ex situ was performed and operando dataset were found reliable for cell parameters and zO position refinement. both both datasets, due to the strong neutron cross section of Ni compared to Li, neutron diffraction pattern changes only slightly with Li occupancy, which was hence not refined.

### X-ray absorption spectroscopy (XAS)

XAS experiments were carried out at ROCK [61] beamline, SOLEIL synchrotron. A toroidal Ir coated mirror focalizes in the horizontal plan the photons emitted by a 3T super bend magnet. The harmonic rejection is provided by two mirrors tilted at 3 mrad using the B4C stripes. The Ni K edge measurements were performed by using the Si(111) quick-xas monochromator. The oscillation frequency of 2Hz was used. The signal was collected in transmission mode using three gas ionization chambers in series as detectors, allowing the simultaneous recording of a Ni metal foil as reference to calibrate the energy of each spectrum. The beam size was 2.5 mm X 0.650 mm (HXV) at the sample position. The XAS spectra were collected in at least 2 positions, to ensure the repeatability of the measurement and assess on the beam damage. A new multimodal electrochemical cell was used for the operando experiments, a complete description of the cell is provided in the ref. The cell assembled using Al-supported LNO electrode supplied by BASF, 28 mL of LP57 (Elyte), Li metal, and Celgard as separator. Energy calibration and spectral normalization were carried out using the graphical interfaces available on the ROCK beamline[62]. The spectra are averaged over two minutes, corresponding to a variation of around 3.33 mM of Li at the C/10 rate. A python library ("Perex") to aggregate the quick-XAS and electrochemical data was used (https://github.com/GhostDeini/perex). Quick-XAS data grouped into matrices were analyzed by combining Principal Component Analysis (PCA) and Multivariate Curve Resolution with Alternating Least Squares (MCR-ALS), as already reported in previous studies[63]. The EXAFS of three reconstructed principal components and pristine LNO were modeled in R space using the Demeter package[64]. Different models were considered, and the parameters of interest are shown in the table 1 and table 2 of SI.

### Resonant Inelastic X-ray scattering (RIXS)

The high-energy resolution RIXS measurements were carried out at the ID32 beamline of the ESRF [65]. The incident photon energy was tuned to the $L_3$ absorption edge of Ni and scan across the edge (~850 - 860 eV) while resolving in energy in emitted photons. The energy resolution of the emitted photon was set to ~25 meV with a beam spot of 40 µm×3.5µm. X-ray beam hit the sample perpendicularly and the emitted photons are also measured almost perpendicularly to the sample surface. Cycled and pristine LNO and NNO electrodes were glued on electronically conducting sample holder using Ag paint in a glovebox. Sample transfer between the glovebox and the high vacuum measurement chamber ($10^{-9}$ mbar) was performed without air-exposure using $N_2$ glove bag. RIXS have been measured at 25 K and RT. RIXS spectra have long acquisition times – typically on the order of 30 min for one excitation energy. No evolution of RIXS spectra was observed as a function of measurement time (beam damage doesn't influence RIXS spectra). Data analysis was performed using the RIXStoolbox[66]. Ni L-edge simulations were performed using Quanty code and available parameters in literature[67]

**Raman spectroscopy**

Pristine LiNiO2 and NaNiO2 powders were measured under inert atmosphere in the EL-CELL. LiNiO2 cathode pellets were prepared inside the Ar-filled glovebox using 80% LiNiO2, 10% conductive carbon Super P and 10% PTFE as a binder. Freestanding LNO films were dried in Buchi oven at 80°C overnight. Operando Raman measurements were performed in the commercially available EL-CELL using Li metal as counter electrode, LP57 electrolyte (Elyte) and Celgard 2500 separator. Galvanostatic cycling at C/8.3 (based on capacity of 225 mAh/g) was performed with Gamry potentiostat. Raman measurements were performed on HORIBA LabRam HR Evolution confocal Raman spectrometer using 50x long focus lens magnification, 633 nm laser wavelength, 300 grooves/mm grating and 5 minutes total acquisition time. Background subtraction and peak fitting with Lorentzian profile of the spectra has been performed using phyton-based processing software PRISMA[68].

**Hard X-ray photoelectron spectroscopy (HAXPES)**

Ex situ HAXPES spectra were recorded at GALAXIES beamline[69], SOLEIL synchrotron (France). Al-coated LNO electrode was cycled in a coin cell using Li metal as counter electrode, LP57 electrolyte (Elyte) and Celgard 2500 as separator. Galvanostatic cycling at C/4 (based on capacity of 225 mAh/g) was applied, then the electrode was recovered and rinsed by DMC, to avoid electrolyte contamination on the electrode surface. NaNiO2 polycrystalline powder was sputtered on carbon tape and placed on the same Omicron plate of LNO samples. A vacuum chamber was used to transfer the samples from the argon glovebox to the HAXPES introduction chamber to avoid air exposure. Photon excitation energies of hν = 3.7 keV and 8.6 keV were obtained from the first and the third order reflections of the Si(111) double-crystal monochromator, respectively. The photoelectrons were collected by a SCIENTA EW4000 analyzer, and the obtained energy resolution from the Au Fermi edge were 0.4 eV for 3.7 keV and 0.15 eV for 8.6 keV. The probing depths of 14 and 33 nm was determined by the TPP-2M model developed by Tanuma *et al*.[42], at 3.7 keV and 8.6 keV incident photon energies, respectively. The analysis chamber pressure was always kept at around $10^{-8}$ mbar during the whole experiment, and no charge neutralizer was required. The binding energy scale was calibrated with the C 1s component at 290.9 eV attributed to $CF_2$ group of PVdF binder present in the samples.

**X-ray Raman scattering (XRS)**

All the XRS data collected at the beamline ID20 of the ESRF [70]. The pink beam from four U26 undulators was monochromatized to an incident energy of 9.6837 keV, using a cryogenically cooled

Si(111) monochromator and focused to a spot size of approximately 50 micron by 20 micron (VxH) at the sample position using a mirror system in Kirkpatrick–Baez geometry. The large solid angle spectrometer at ID20 was used to collect XRS data with 36 spherically bent Si(660) analyzer crystals. The data were treated with the XRStools program package as described elsewhere[71]. For LNO, we have used the electrodes supplied by BASF, while a pellet was made for NaNiO2. The samples were placed into the beam so to have a 10-degree grazing incident beam. All XRS measurements were collected at room temperature. The signals were collected with a 0.2 eV (close to the edge) or 0.7 eV (far from the edge) step by scanning the incident beam energy to record energy losses in the vicinity of the O K-edge (520–565 eV) and Ni $L_{2,3}$-edge (844-878 eV). Acquisition scans lasted around 3–4 h per sample. All scans were checked for consistency before averaging over them. The overall energy resolution of the XRS spectra was 0.7 eV as estimated from the FWHM of elastic scattering from a piece of adhesive tape. Also, signals from analyzer crystals at different scattering angles were measured, covering a momentum transfer from 2.5 to 9.2 Å$^{-1}$. The data were integrated at medium and high q for O and Ni XRS spectra, respectively.

**XSpectra calculation of the XANES spectra**

Ab initio XANES spectra for the Ni (K and L2,3-edge) and O (K-edge) were calculated using the XSpectra module [25,26] of the Quantum ESPRESSO package [72], which implements a plane-wave, pseudopotential approach based on density functional theory (DFT). All structures used in DFT calculations were obtained with permission from previously published works [13]. All calculations were performed using Monkhorst-Pack [73] K-point meshes generated for each system based on a K-point distance of 0.25 Å-1 and utilised a PBE+U+J approach [74] with U = 5.0 and J = 0.5 eV based on values from a previous study [75]. Calculations of XANES spectra used a final-state approximation with a localised core-hole, for which supercells of each structure were generated in order to maintain a minimum separation distance between neighbouring images of 8 Å in all cases. To correctly align the spectra from all different absorbing atoms onto a common relative energy scale, an energy-alignment procedure based on the delta-KS method for computing core-level binding energies was used for all individual spectra components unless otherwise stated[76].

**FDMNES calculations of the O XRS spectra**

Ab initio simulations were conducted using the Finite Difference Method Near Edge Structure (FDMNES) software [77,78]. The generation of atomic positions from space group n. 166:H for both pristine and delithiated material (pristine: a=b=2.8831, c=14.1991 alpha=beta=90, gamma=120; delithiated: a=b=2.83527, c=14.3319 alpha=beta=90, gamma=120;), with appropriate atomic occupancies, was the input parameters the simulations. Employing multiple scattering theory based on the muffin-tin (MT) approximation for the potential shape of the Green scheme, we utilized non-relativistic calculations and the Hedin-Lundqvist exchange. Tuning MT radii ensured a good overlap between different spherical potentials. The resulting spectra were convoluted and normalized to the same integrated intensity as the experimental data in the appropriate energy loss range. Additionally, we applied semiempirical parameters, screening and dilatorb, to consider various interactions between Ni d and O 2p orbitals (where smaller screening indicates stronger electronic correlation between the scattering atom and its environment, typically associated with larger hybridization and correspondingly lower oxide ion basicity). We also considered potential increased orbital overlap (higher dilatorb, accounting for larger hybridization due to increased atomic orbital overlap). The increase in the dilatorb value primarily induces variation in the pre-edge peak intensity and a slight shift to lower energy, but it also results in a decrease in the intensity of peak (c). Notably, changes in the screening parameter, reflecting to some extent the interaction between oxygen and Ni d states, lead to significant alterations in the intensity ratio and mutual position of the O main edge and pre-edge peaks. An increase in the screening value causes a drastic shift of the pre-edge and main peak to higher energies. The utilization of different combined dilatorb and screening values minimizes the discrepancy between experimental and calculated spectra; optimal results for the pristine are obtained for dilatorb 0.2 and screening 0.9, whereas for the delithiated sample are dilatorb 0.1 and screening 0.9.

## Acknowledgements

This project has received funding from the European Union's Horizon 2020 research and innovation programme under grant agreement No 957189. Beamtime at the ESRF on ID32 and ID20 was granted within the Battery Pilot Hub MA4929 "Multi-scale Multi-techniques investigations of Li-ion batteries: towards a European Battery Hub". SOLEIL synchrotron is acknowledged for providing in-house access to ROCK and GALAXIES beamlines (20210715, 99230154). ILL (institute Laue-Langevin) is also acknowledged for beamtime on D2B (EASY-886) and D19 ([DIR-221](DIR-221)). High-performance computing resources and support have been granted by CINECA award under the ISCRA initiative. This work was supported by a public grant overseen by the French National Research Agency (ANR) as part of the "Investissements d'Avenir" program (reference: ANR-10-EQPX-45). ANR is also acknowledged for funding through the STORE-EX Labex Project ANR-10-LABX-76-01. We thank Sabrina Sicolo (BASF) and Alessandro Mirone (ESRF) for their support in theoretical calculations and fruitful discussions. Elisa Grepin and Sathya Mariyappan (Collège de France) are warmly thank for the synthesis of $NaNiO_2$.

## Author contributions

Q. Jacquet, S. Lyonnard, A. Iadecola conceived the idea. S. Lyonnard managed the project. N. Mozhzhukhina, A. Matic performed the Raman measurement. G. Wittman, L. R. Perez, S. Belin, A. Iadecola performed the XAS measurements. P. O. Gillespie, D. Prezzi performed the XAS and XRS calculations. A. Iadecola, F. Capone, J.P. Rueff, R. Dedryvère, performed the HAXPES measurements. A. Longo, L. Stievano, A. Iadecola, Q. Jacquet and S. Lyonnard performed the XRS measurements. A. Longo performed the XRS calculations. Q. Jacquet, E. Suard performed the neutron diffraction measurements. Q. Jacquet, N. B. Brookes performed the RIXS measurements. Q. Jacquet, A. Iadecola, N. Mozhzhukhina, P. Gillespie, R. Dedryvère, D. Prezzi wrote the manuscript. All authors gave feedback and participated to the discussion.

# Supporting information

## Neutron diffraction

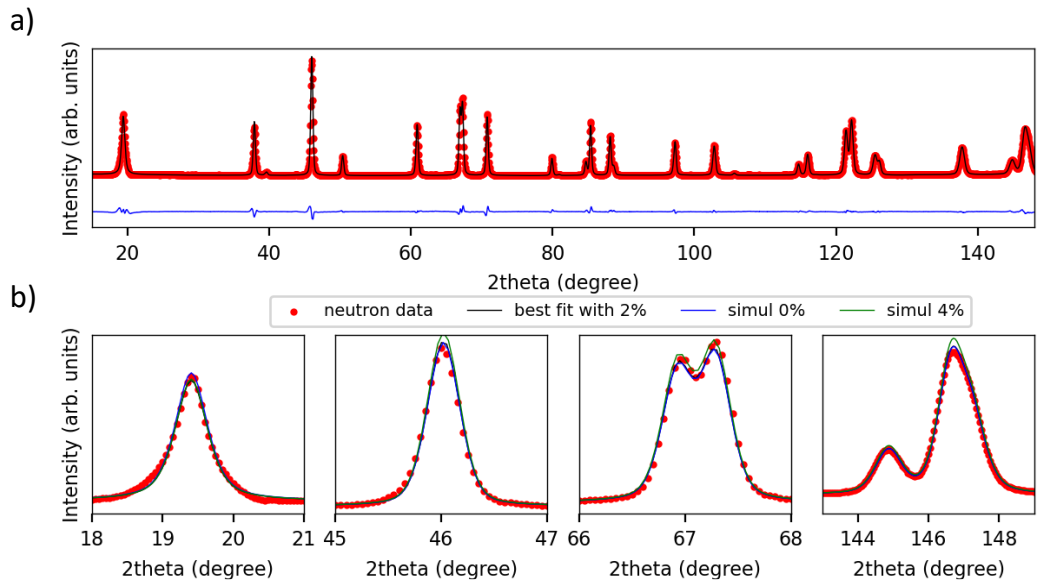

**Figure S1**: a) Ex situ high resolution neutron diffraction and Rietveld refinement of pristine LiNiO$_2$ powder R$_{Bragg}$ = 2.13. red dots, black line and blue line are the data, simulated patterns and difference, respectively. b) four zooms for the Rietveld refinement shown in a) with additionally simulated patterns corresponding to different fraction of Ni anti-site mixing namely 0 and 4 % in blue and green respectively. Note that the influence of Ni anti-site mixing on the pattern is weak, hence, the refinement achievable only for high quality data and fit.

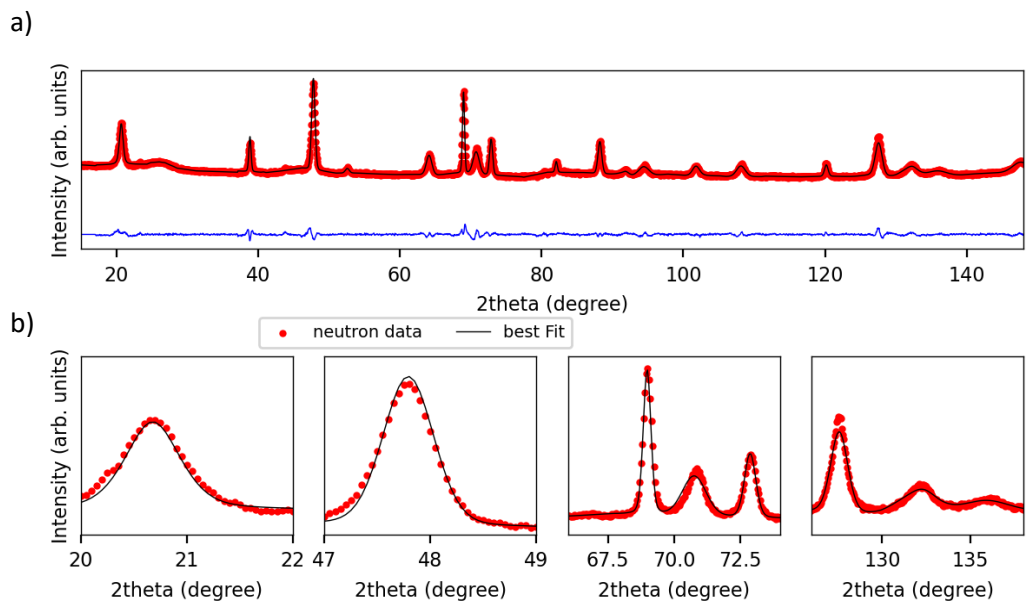

**Figure S2**: a) Ex situ high resolution neutron diffraction and Rietveld refinement of Li$_x$NiO$_2$ charged at 4.5 V, R$_{Bragg}$= 3.06. Red dots, black line and blue line are the data, simulated patterns and difference, respectively. b) four zooms for the Rietveld refinement shown in a).

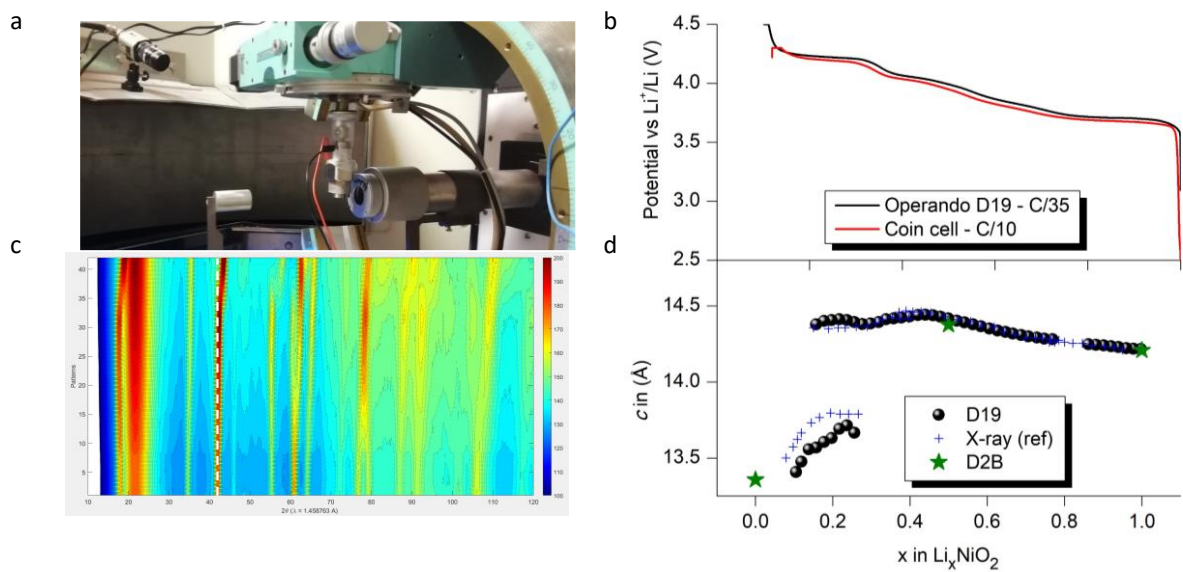

**Figure S3**: Operando neutron diffraction experiment on LNO on D19 (ILL). a) picture of the cell in the instrument. b) electrochemical profile of the operando (black) versus coin cell (red) for LNO half-cell. c) neutron diffraction pattern evolution. d) c lattice parameter obtained with Rietveld refinement of the operando neutron powder diffraction data compared to ex situ neutron powder diffraction measurement and x-ray from literature [ref De Biasi].

**Operando XAS**

The operando XAS dataset contained 204 spectra recorded at two different positions on the cathode. The initial guess of the spectra in the ALS minimization was done using the SIMPLISMA method. Constraints have been applied at concentration matrix: non-negativity, unimodality (since we have decomposed the matrix into four sub-matrices corresponding to the first position in charge then in discharge, then to the second position in charge then in discharge), closure at 1 and the concentration of the first component was fixed at 100% for the first data; for the matrices of spectra we have imposed only non -negativity.

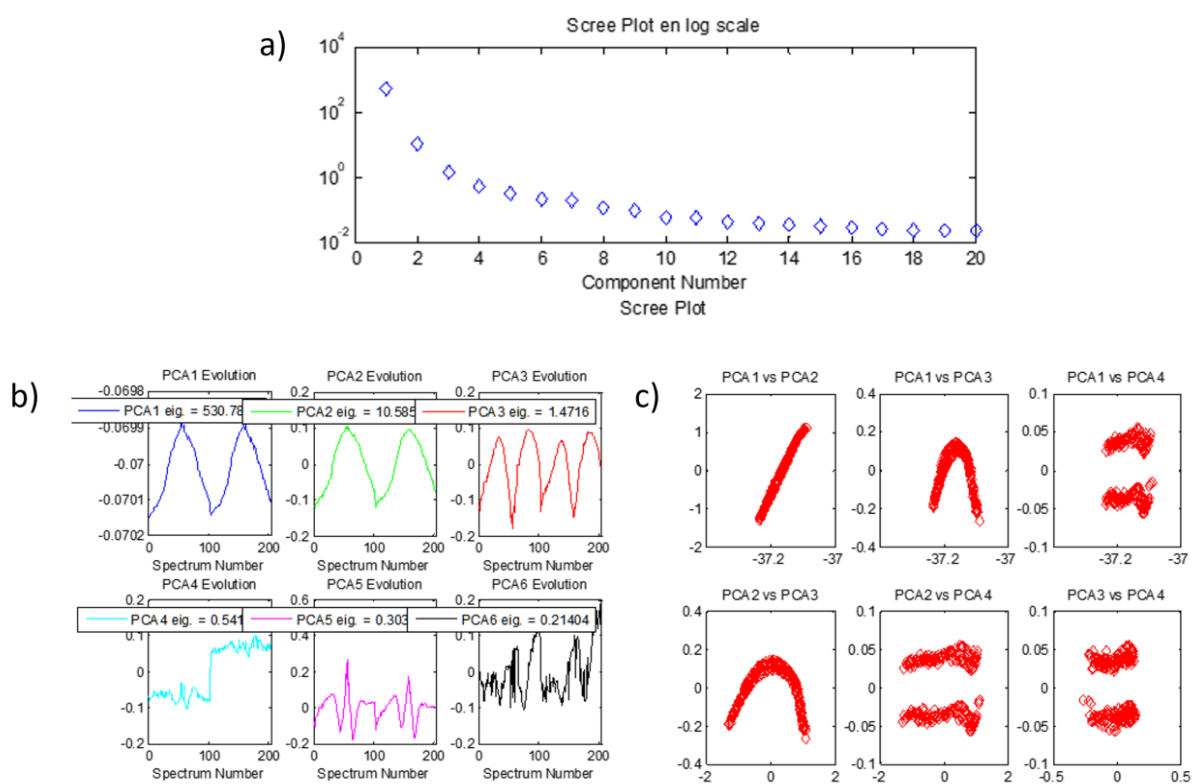

**Figure S4**: a) Screen Plot, b) Evolution of the component scores during the first cycle, c) Projection of scores in space.

**Table S1** Radial distances and corresponding mean-square relative displacements obtained from the EXAFS refinement of the pristine LNO electrode and NNO powder. The $k^2$-weighted EXAFS oscillations of the reconstructed principal components were extracted using the sine function in the k-range [2.7-17.7] Å$^{-1}$ and they were modeled in the R space from 1 to 3.0 Å. The $E_0$ for the background extraction was fixed at 8342 eV and $S_0^2$ at 0.85 for the EXAFS refinement.

| Species | Space group | Shell | E0 (eV) | Dist. (Å) | C.N. | σ (Å$^{-2}$) | R-factor |
|---|---|---|---|---|---|---|---|
| LNO | C2/m | Ni-O | -4.922 | 1.92 | 4 | 0.00455 | 0.00645 |
| | | Ni-O | | 2.04 | 2 | 0.00455 | |
| | | Ni-Ni | | 2.88 | 6 | 0.00428 | |
| LNO | R-3C | Ni-O | -5.448 | 1.95 | 6 | 0.00816 | 0.00959 |
| | | Ni-Ni | | 2.88 | 6 | 0.00473 | |
| LNO | | Ni-O | -4.536 | 1.87 | 2 | 0.00465 | 0.00379 |
| | | Ni-O | | 1.92 | 2 | 0.00465 | |
| | | Ni-O | | 2.04 | 2 | 0.00465 | |
| | | Ni-Ni | | 2.88 | 6 | 0.00433 | |
| NaNiO2 | C2/m | Ni-O | -1.159 | 1.91 | 4 | 0.00457 | 0.00992 |
| | | Ni-O | | 2.14 | 2 | 0.00681 | |
| | | Ni-Ni | | 2.84 | 2 | 0.00409 | |
| | | Ni-Ni | | 3.01 | 4 | 0.00683 | |

**Table S2** Radial distances and corresponding mean-square relative displacements obtained from the EXAFS refinement of the reconstructed principal components. The $k^2$-weighted EXAFS oscillations of the reconstructed principal components were extracted using the Kaisser-Bessel function in the k-range [3-13.2] Å$^{-1}$ and they were modeled in the R space from 1 to 2.9 Å. The $E_0$ for the background extraction was fixed at 8342 eV and $S_0^2$ at 0.85 for the EXAFS refinement.

| Species | Space group | Shell | E0 (eV) | Dist. (Å) | C.N. | σ (Å$^{-2}$) | R-factor |
|---|---|---|---|---|---|---|---|
| PC 1 | R-3C | Ni-O | 1.753 | 1.955 | 6 | 0.01165 | 0.00327 |
|      |      | Ni-Ni |      | 2.89  | 6 | 0.00471 |         |
| PC 1 | C2/m | Ni-O | 1.933 | 1.91  | 1.75 | 0.00592 | 0.00389 |
|      |      | Ni-O |       | 1.99  | 4.25 | 0.01192 |         |
|      |      | Ni-Ni |      | 2.89  | 6    | 0.00464 |         |
| PC 2 | R-3C | Ni-O | 0.966 | 1.90  | 6 | 0.00816 | 0.00762 |
|      |      | Ni-Ni |      | 2.855 | 6 | 0.00473 |         |
| PC 2 | C2/m | Ni-O | 1.894 | 1.89  | 4.76 | 0.00478 | 0.00479 |
|      |      | Ni-O |       | 2.06  | 1.24 | 0.01304 |         |
|      |      | Ni-Ni |      | 2.86  | 6    | 0.00467 |         |
| PC 3 | R-3C | Ni-O | 1.055 | 1.88  | 6 | 0.00564 | 0.00854 |
|      |      | Ni-Ni |      | 2.83  | 6 | 0.00414 |         |
| PC 3 | C2/m | Ni-O | 0.799 | 1.905 | 4.43 | 0.00262 | 0.004456 |
|      |      | Ni-O |       | 1.80  | 1.57 | 0.00137 |         |
|      |      | Ni-Ni |      | 2.82  | 6    | 0.00405 |         |
| PC 3 | C2/m | Ni-O | 1.061 | 1.88  | 6 | 0.00498 | 0.00504 |
|      |      | Ni-Ni |      | 2.82  | 6 | 0.00407 |         |

**Ex situ Raman**

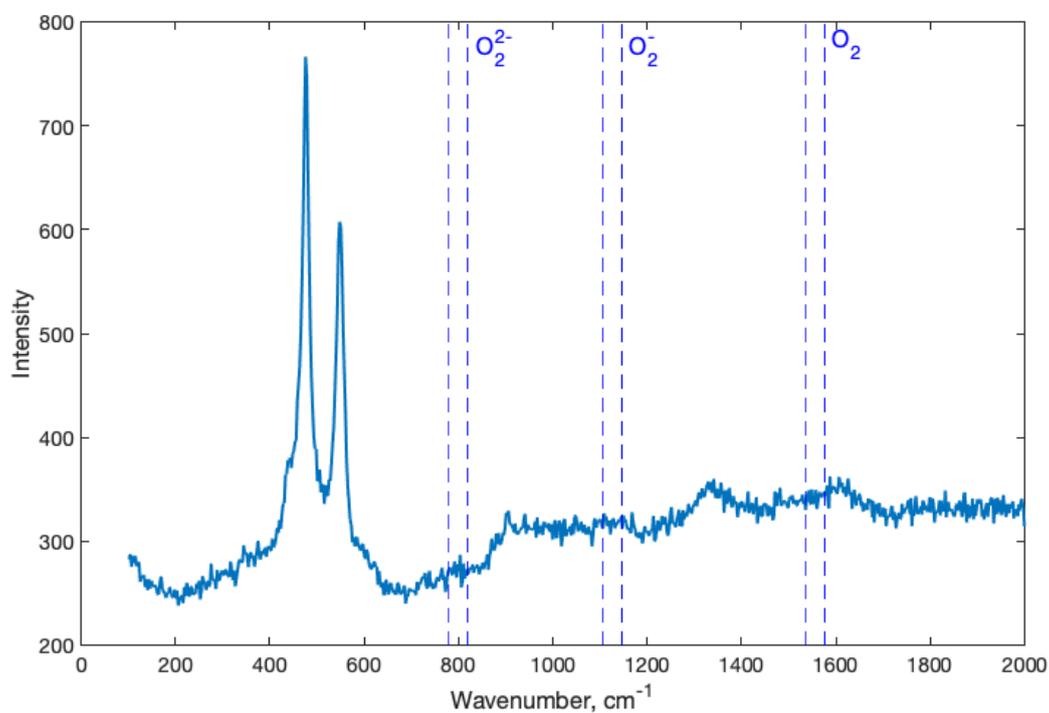

**Figure S5**: Extended Raman spectrum of LNO charged to 4.3 V. Raman wavenumber intervals for expected peroxide, superoxide and molecular oxygen species are highlighted.

**Ex situ HAXPES**

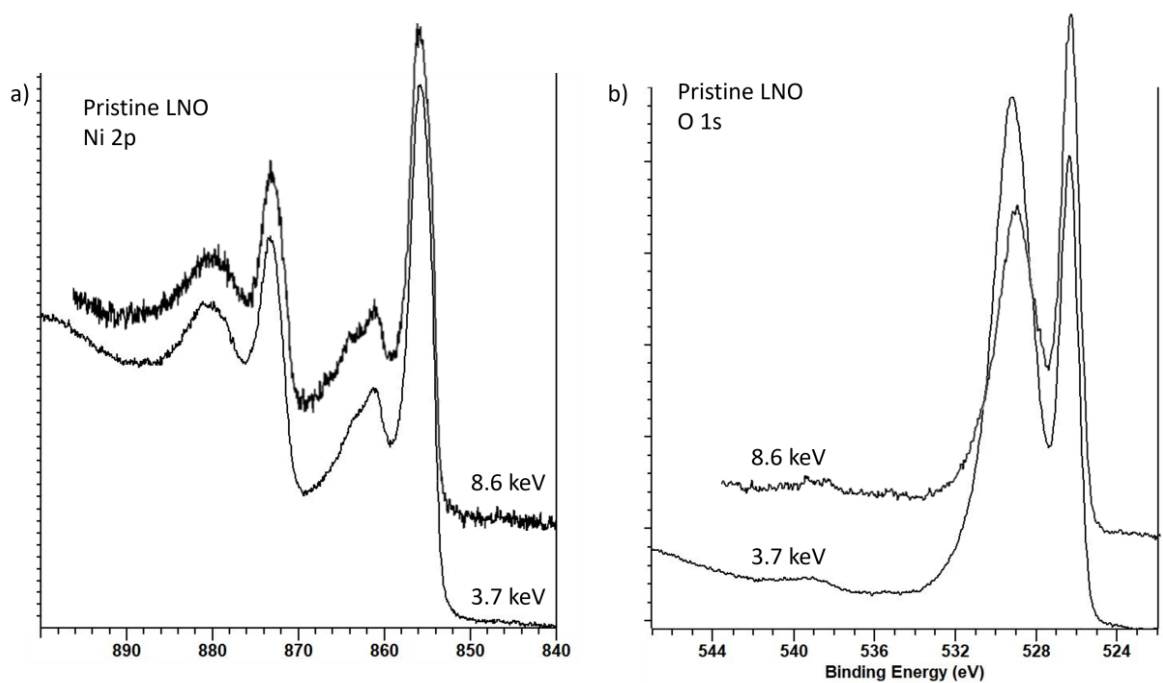

**Figure S6**: a) Ni 2p and b) O 1s ex situ HAXPES spectra obtained at different incident energy on LNO pristine electrode.

**Ex situ XRS**

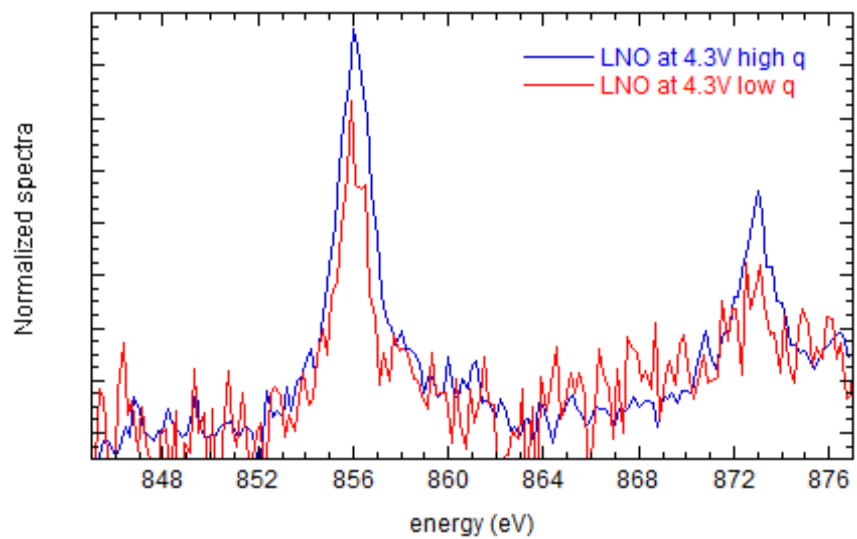

**Figure S7**: XRS spectra obtained at Ni L2,3-edge in the charged LNO electrode at low and high q.

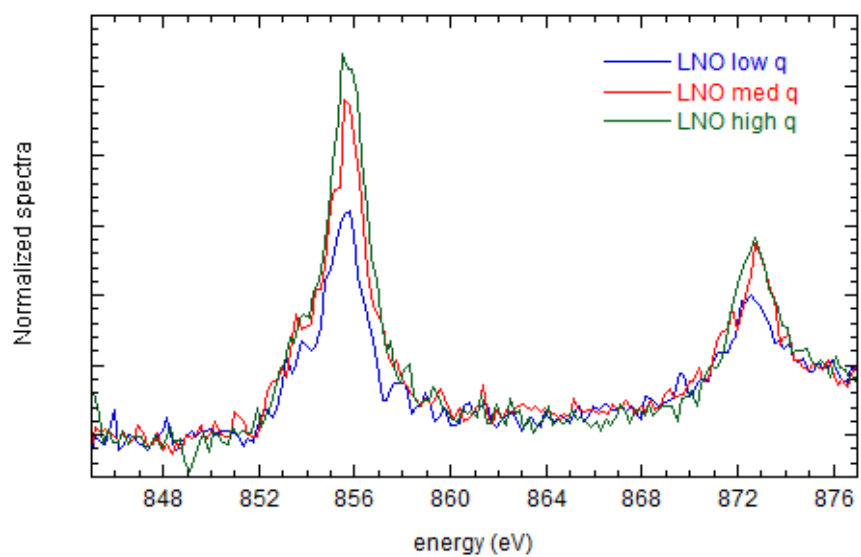

**Figure S8**: XRS spectra obtained at Ni L2,3-edge in the pristine LNO electrode at different q.

**Discussion on Ni K-edge XANES simulations**

The final Ni K-edge XANES spectra (Fig.2, main paper) are computed with the XSpectra module [Calandra2009, Bunău2013] of the Quantum ESPRESSO package [Gianozzi2020, Gianozzi2017, Gianozzi2009]. The spectra are obtained by applying the Lanczos algorithm, with a minimal 0.5 eV constant Lorentzian broadening (subplot C, Fig. S9). Variable-energy broadening, using the approach outlined by Bunău and Calandra [Bunău2013] and based on the Seah-Dench formalism for the inelastic mean-free-path [Seah1979], is then applied as a post-processing to include lifetime effects and ease the comparison to experiments. For the representation of pristine $LiNiO_2$ we considered both the 100% JT ($P2_1/c$) and 100% BD ($P2/c$) models as well as intermediate compositions, proposed as being almost energetically degenerate with the established JT structure [Foyevtsova2019]. We find in this instance that the Ni K-edge spectra are nearly identical when lifetime effects are included.

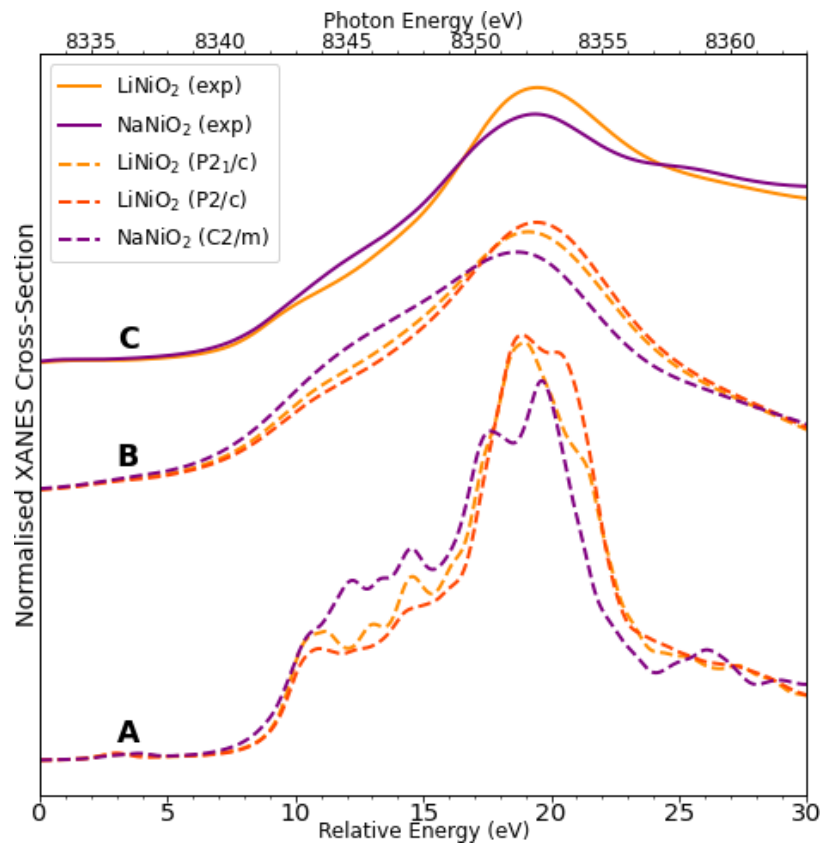

**Figure S9**: Ni K-edge XANES spectra of $P2_1/c$ and $P2/c$ $LiNiO_2$ (red-brown and orange, respectively) and $NaNiO_2$ (purple) simulated with XSpectra (subplots **A&B**) compared to experimental data (subplot **C**) for $LiNiO_2$ (orange) and $NaNiO_2$ (purple) from this work. Computed spectra in subplot **B** have been post-processed with a variable-energy broadening scheme reported previously in the literature [Bunău2013]. To better show the underlying spectral features, we show the same spectra with a constant 0.5 eV Lorentzian broadening (subplot **A**).

## Discussion of Ni environment & Charge State

In order to better understand the chemical environments of different Ni sites seen in our structures in relation to their XANES spectra, we conducted an analysis of the electronic structures and formal oxidation states of Ni and O in the pristine LNO structures used here ($P2_1/c$ & $P2/c$), as well as $NiO_2$ and $NaNiO_2$ (Table S3). The analysis considered the Bader charge in the unit cell as well as the spin-polarisation recorded by Löwdin charge analysis by Quantum ESPRESSO. To further interrogate the d-orbital occupations of the Ni atoms themselves, we also employed a method defined previously by Selloni *et al.* [Sit2011] in which the formal oxidation state (ionic character) may be determined by counting the d-orbital occupations reported by Quantum ESPRESSO in the electronic structure calculation. For this, we determine a formal oxidation state for each Ni atom using a minimum threshold for formal occupancy of 0.99 for the d-orbital occupation number in all cases.

**Table S3**: Tabulated Bader charge analysis of Ni and O sites from different structures used in this study.

| System | Element | Site # | Bader Charge | Spin Polarisation (bohr mag) | Ni formal oxidation state |
|---|---|---|---|---|---|
| $LiNiO_2$ ($P2_1/c$) | Ni | 3, 4 | +1.28 | 0.92 | 3+ |
|  | O | 5 - 9 | -1.09 | 0.03 | - |
| $LiNiO_2$ ($P2/c$) | Ni | 5, 6 | +1.19 | 1.62 | 2+ |
|  | Ni | 7, 8 | +1.33 | 0.10 | 4+ |
|  | O | 9 - 16 | -1.02 / -1.13[1] | 0.03 | - |
| $NaNiO_2$ | Ni | 2 | 1.30 | 0.98 | 3+ |
|  | O | 3 | -1.07 | 0.01 | - |
| $NiO_2$ | Ni | 1 | +1.33 | 0.00 | 4+ |
|  | O | 2 | -0.67 | 0.03 | 4+ |

[1] Values from two different oxygen sublattices.

From the analysis we observe comparable results of Bader charge analysis performed on systems with a similar lithium/sodium content, in agreement with a recent study on the same subject [Genereith-Schriever2023]. Combining the formal oxidation state determined by the method of Selloni *et al.* with the reported spin-polarisation, we observe formal oxidation states for the P2/c structure of Ni2+ for two of the sites and Ni4+ in the remaining two, in accordance with symmetry analysis. For the P2$_1$/c structure of LiNiO$_2$ and NaNiO$_2$, we find a formal oxidation state of Ni3+ for all Ni atoms. For NiO$_2$, we find a formal oxidation state of Ni4+.

We note that, despite there being very little difference in Bader charge between different Ni environments, the recorded spin state of each site appears to correspond to different formal oxidation states of Ni. This deviation between formal charges based on a partition-of-space analysis (of the sort employed in the Bader method [Bader1990]) and d-orbital occupancy in transition metal oxides has been noted in previous works [Timrov2022, Timrov2023, Raebiger2008, Sit2011]. In particular the work of Raebiger *et al.* [Raebiger2008] has proposed a charge-compensation mechanism [Resta2008, Jansen2008] which accounts for the lack of change in the net physical charge of the TM ion due to changes in the ligand environment.

Considering the computed oxidation states (Table S3), we observe a difference between the computed Ni K-edge spectra arising from sites containing Ni in different oxidation states (Fig. S10), particularly between the two Ni sites of the P2/c structure, which naturally follows from each site having different local bonding environments.

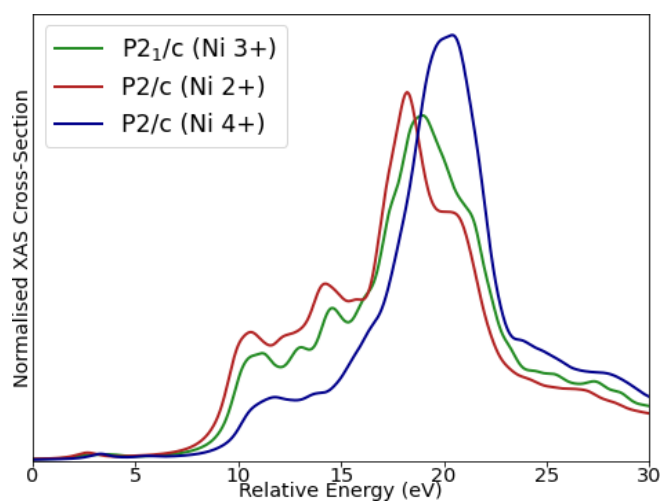

**Figure S10**: Ni K-edge XANES spectra for different Ni sites of P2$_1$/c LiNiO$_2$ (green) and of P2/c LiNiO$_2$ (red and blue). All spectra shown are plotted with a constant 0.5 eV Lorentzian broadening scheme to more clearly show differences between different peaks.

## XRS simulation using FDMNES

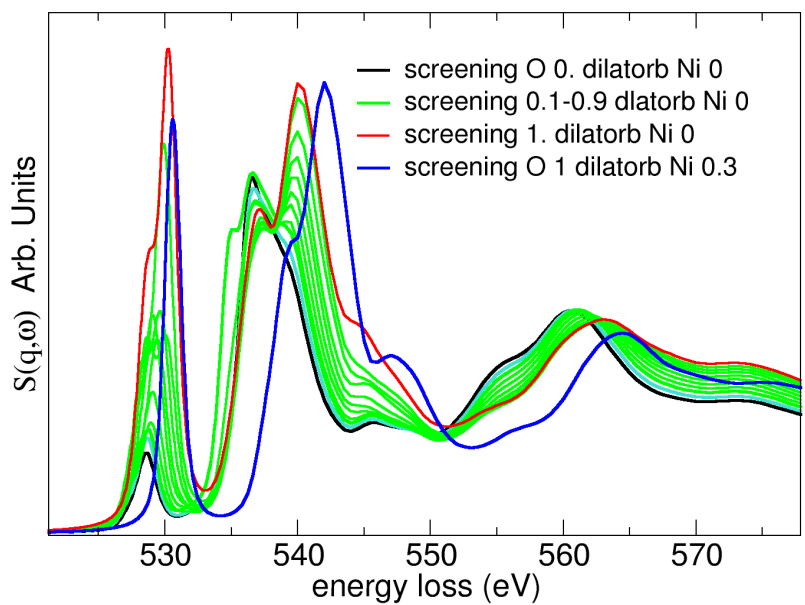

**Figure S11** Simulated O K-edge XRS of pristine LiNiO$_2$ using FDMNES and varying the screening and dilatorb parameters.